\documentclass[a4paper,11pt]{article}
\pdfoutput=1 % if your are submitting a pdflatex (i.e. if you have
\usepackage{jinstpub} % for details on the use of the package, please
\usepackage{mhchem}
\usepackage{siunitx}
\title{The upgrade of the CMS muon system for the High Luminosity LHC}

\collaboration[c]{on behalf of the CMS Muon group}
\author[a]{Antonello Pellecchia}

\affiliation[a]{INFN, sezione di Bari, Italy}

\emailAdd{antonello.pellecchia@ba.infn.it}

\abstract{
    The muon system of the CMS experiment is expected to upgrade all of its subdetectors for the Phase-2 of the Large Hadron Collider (LHC) that will begin in 2029.
    The upgrade plans for drift tubes (DTs), cathode strip chambers (CSCs) and resistive place chambers (RPCs) include a new electronics for better performance in high background rate conditions and to sustain the large radiation dose delivered by the high-luminosity LHC (HL-LHC).
    Two new RPC stations will also be installed to complement CSCs in the forward region $1.8 < |\eta| < 2.4$, while three stations of triple-GEM (gas electron multiplier) detectors will complement CSCs in the region $1.5 < |\eta| < 2.4$ and extend the CMS muon system coverage to the very forward pseudorapidity region $2.4 < |\eta| < 2.8$.
    We present the goals and status of the CMS muon system upgrade including the performance of the early Phase-2 upgrades demonstrated in Run 3 with proton-proton collisions, the performance validation of the new detectors in test beams and the mass production status of the new stations.
	}


\collaboration[c]{on behalf of the CMS Muon group}

\proceeding{
	IPRD2023: 16$^{\text{th}}$ Topical Seminar on Innovative Particle and Radiation Detectors\\
	25-29 Sep 2023\\
	Siena (Italy)
	}

\begin{document}
\maketitle
\flushbottom

\section{Introduction}
During the Run 2 of the Large Hardon Collider (LHC), the CMS experiment \cite{ref:cms} collected 160~fb$^{-1}$ of proton-proton collision data at 13 TeV, while the Run 3, started in 2021, increased the LHC collision energy to 13.6 TeV.
The Run 3 concludes the Phase-1 of LHC, while Phase-2 will consist of a High-Luminosity upgrade of LHC (HL-LHC); HL-LHC will deliver an instantaneous luminosity up to \SI{7.5}{\per\centi\m\squared\per\second} at 14 TeV, for an integrated luminosity up to 4000~fb$^{-1}$.
The CMS experiment will be upgraded in all of its subdetectors to maintain a good trigger and offline event reconstruction performance under the higher collision rates, increasing its Level-1 trigger rate from 100 to \SI{750}{\kilo\Hz} and its latency buffer to \SI{12.5}{\micro\s}.

\begin{figure}[tb]
    \centering
    \includegraphics[width=.8\textwidth]{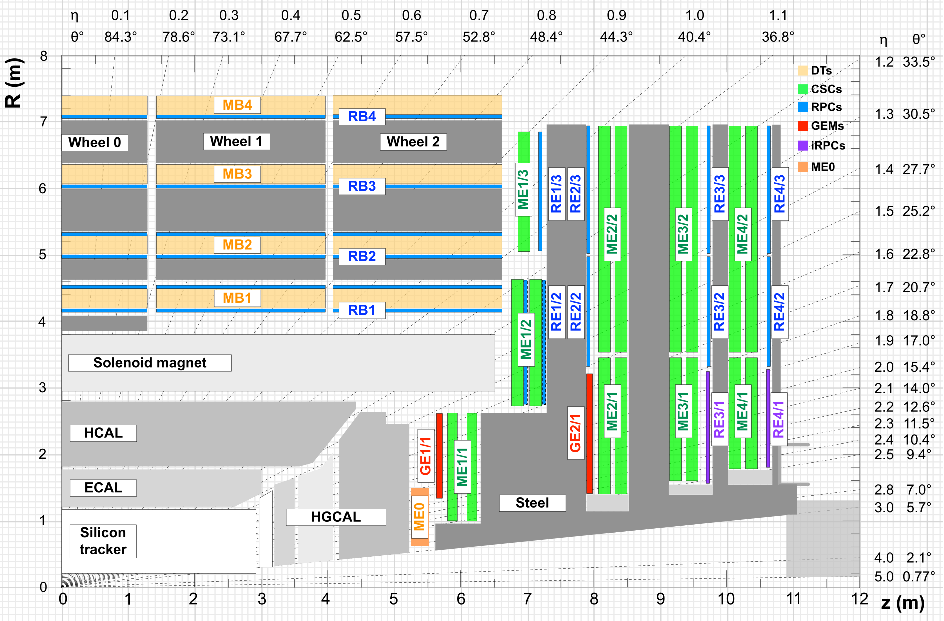}
    \caption{Quadrant of a section of the CMS muon spectrometer, showing the Phase-1 muon stations and the new stations after the Phase-2 upgrade (in red, violet and orange) \cite{ref:muon_tdr}.}
    \label{fig:cms_muons}
\end{figure}

The CMS muon spectrometer will upgrade its subdetectors to sustain the higher trigger rates and background particle rates \cite{ref:muon_tdr}.
The Phase-1 CMS muon system is instrumented with three gaseous detector technologies:
drift tubes (DTs) in the barrel, cathode strip chambers (CSCs) in the endcaps and resistive plate chambers (RPCs) in both regions.
In the Phase-2 upgrade (Fig.\,\ref{fig:cms_muons}), parts of the legacy electronics are being replaced with components able to handle the higher doses in the CMS cavern, ensure higher data bandwidth and deliver better performance for background rejection;
all three detector technologies are also investigating the detector longevity under irradiation with doses higher than the expected ones in HL-LHC.
The upgrade also includes the installation of two new RPC stations (RE3/1 and RE4/1) to complement CSCs in the muon forward region and three new stations of triple-GEM (gas electron multiplier) detectors (GE1/1, GE2/1 and ME0) to complement the CSC systems and extend the muon system coverage in the very forward region.

The first results of the early Phase-2 upgrades completed in the LHC Long Shutdown 2 (LS2, which lasted from 2019 to 2021) have been demonstrated with CMS cosmic runs and the Run 3 collision data.
Mass production for all the new detector stations has started or is expected to start by 2024.

\section{DT upgrade}

The barrel of the CMS muon system is fully covered by drift tubes, consisting of multiple cells of single-wire counters arranged in 4 concentric ring stations -- in turn divided in five wheels.
The DT system provides an efficiency larger than 99\%, a space resolution of \SI{100}{\micro\m} for muon segments and a time resolution of 2-\SI{3}{\nano\s}.

During the HL-LHC operations, a degradation of the DT performance over time is expected due to ageing induced by the background radiation \cite{ref:dt_ageing}.
The ageing has been quantified in irradiation studies performed since 2017 at the CERN Gamma Irradiation Facility (GIF++) \cite{ref:gif}, with visible gain drop for a group of wires up to 90\% at an integrated charge of \SI{40}{\milli\coulomb/\centi\m\squared}, corresponding to 3 times the maximum expected HL-LHC dose.
However, this performance drop is not expected to have a visible impact on the CMS muon reconstruction, also thanks to the redundancy with the RPC stations and the overlap with the CSCs in the highest-pseudorapidity region, where the integrated charge is highest.

On the other hand, the legacy DT readout and trigger electronics require an upgrade due to their inadequate radiation resistance for HL-LHC conditions.
The Phase-2 DT upgrade consists of replacing the Phase-2 electronics with the On-Board DT electronics (OBDT) \cite{ref:obdt}:
the OBDT implements a time-to-digital converter (TDC) to send readout data to the back-end, where the trigger primitives are generated exploiting the full DT time resolution.
This enhancement compared to the legacy trigger system, which only used bunch crossing (BX) granularity at the Level-1 trigger, will allow better background rejection and thus better muon momentum resolution.

\subsection{DT slice test performance}

\begin{figure}[tb]
    \centering
    \includegraphics[width=.49\textwidth]{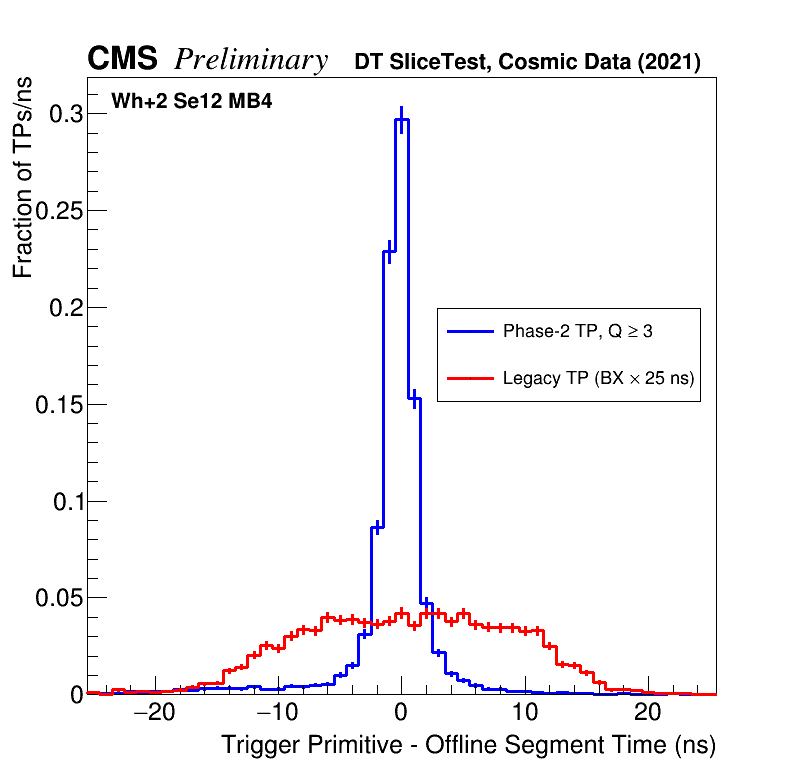}
    \includegraphics[width=.49\textwidth]{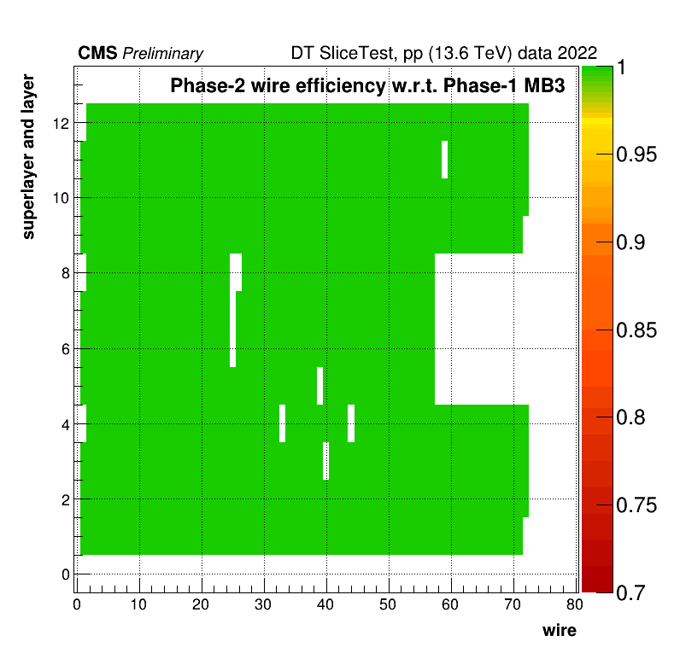}
    \caption{
        (a) Time distribution of the DT trigger primitives with cosmics for the DT slice test using legacy and Phase-2 electronics, showing the better time resolution of the latter \cite{ref:dt_timing}.
        (b) Offline efficiency of the DT slice test measured with respect to the legacy readout in the proton-proton collisions \cite{ref:dt_efficiency}.
        The white areas correspond to non-working readout channels.
    }
    \label{fig:dt_performance}
\end{figure}

A demonstration of the DT upgrade is the slice test, consisting of 13 prototype OBDTs installed during LS2 in one slice of wheel +2.
Throughout the CMS cosmic runs in LS2 and the pp-collision runs of Run 3, the four detectors involved in the slice test have been read out in parallel by legacy and Phase-2 electronics \cite{ref:dt_timing};
Fig.\,\ref{fig:dt_performance}a compares the legacy and Phase-2 muon segment timing measured in cosmic runs, showing the better time resolution of the latter, while Fig.\,\ref{fig:dt_performance}b shows the high efficiency -- larger than 95\%, except for non-communicating readout channels -- of the slice test detectors using the OBDT electronics measured with respect to the legacy readout \cite{ref:dt_efficiency}.
A second slice test has been installed in another sector of the same wheel in 2022 with a second version of the OBDT and has taken data in the 2023 collisions.

\section{CSC upgrades}

The CSC system instruments each of the two muon endcaps with four stations divided in concentric rings;
each slice is an independent module made of six multi-wire proportial chambers with segmented cathodes providing phi position information and anode wires giving eta information.

The legacy CSC front-end electronics consists of the Anode Front End Board (AFEB) and the Cathode Front End Board (CFEB) reading out respectively the anode and cathode signals, complemented by the Anode Local Charged Track (ACLT) board reconstructing muon segments in wire groups.
The off-detector electronics is made of a pair of trigger motherboard (TMB) and data acquisition motherboard (DMB) for each chamber, receiving respectively the trigger and tracking data;
the latter are sent to the back-end front-end driver (FED) for storage \cite{ref:csc_upgrade}.

\subsection{CSC longevity studies}

The CSC detector technology is expected to be able to sustain the increased background rates of the HL-LHC with no observable loss in performance;
this has been pointed out by ageing studies performed at the CERN GIF++ since 2016 \cite{ref:csc_ageing}.
No efficiency decrease was observed up to 10 times the expected Phase-2 integrated charge, although a linear degradation of the space resolution -- within tolerable limits -- has been measured with increasing background rate.
Ongoing performance studies in laboratory and at the GIF++ address the requirement to reduce greenhouse gas consumption.
The gas system of the CSC in CMS -- using 40\% Ar, 50\% \ce{CO2} and 10\% \ce{CF4} -- has been improved for a 60\% \ce{CF4} recirculation, as opposed to the 30\% of Run 2;
measurements with reduced \ce{CF4} concentration (down to 2 and 0\%) and higher \ce{CO2} concentration on a CSC prototype have also shown no gain degradation up to an integrated charge equal to 2.3 times the expected ME2/1 charge at HL-LHC, but carbon deposits are visible on the wires after the irradiation.
Studies with gases alternative to \ce{CF4}, for example HFO1234ze, are ongoing.

\subsection{CSC electronics upgrade}

While the impact of the HL-LHC rates on the intrinsic detector performance are expected to be negligible, the higher background rates will result in higher occupancy per readout channels, which would cause dead time in the legacy electronics;
therefore, the CSC upgrade mainly consists of replacing the electronic boards with newer versions.
The CFEB is replaced with a digital CFEB (DCFEB) which includes a Virtex-6 FPGA and upgraded buffers, while the ALCT boards are upgraded with a Spartan-6 FPGA allowing 9 to 12 times larger memory and bandwidth than its legacy version.
Additionally, both boards feature optical communication with the counting crates through the radiation-hard VTRx and VTTx \cite{ref:vtrx} transceivers and their FPGAs support EEPROM-less programming for better reliability against radiation damage.
The DMB is also replaced by an optical board (ODMB) with higher data bandwidth;
the back-end FED, currently based on the VME standard, will be replaced with an ATCA board called X2O, to be also used in the Phase-2 electronics of the GEM system.
A minimum of 9 X2Os will be used to read out and control the entire CSC system.

\begin{figure}[tb]
    \centering
    \includegraphics[width=.9\textwidth]{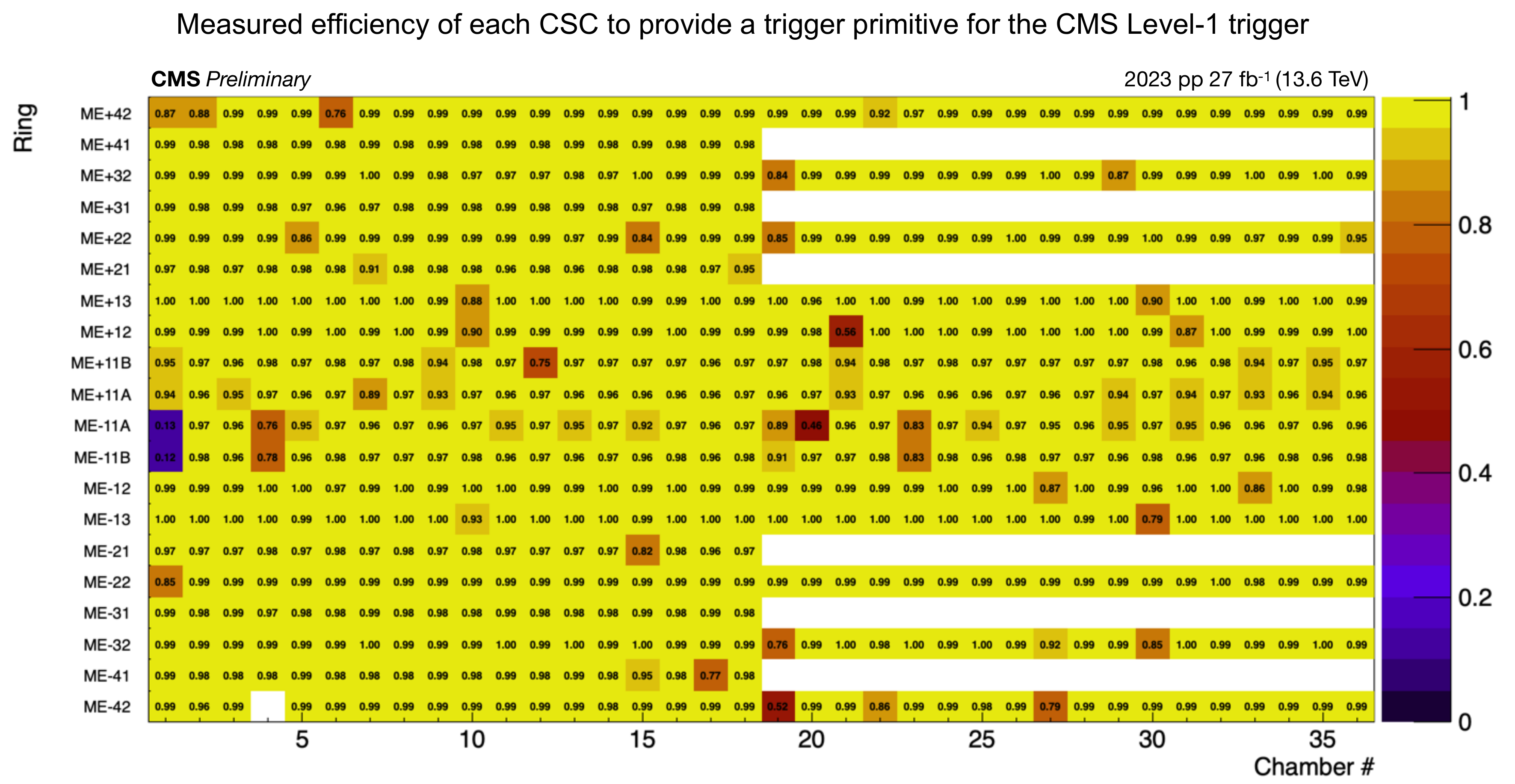}
    \caption{
        Trigger primitive efficiency of each detector of the CSC system measured during the Run 3 pp collisions \cite{ref:csc_efficiency}.
        Chamber 4 of ring ME-42 is permanently disabled because of failed electronics which cannot be accessed without invasive access;
        the other detectors with efficiency lower than 98\% have non-working electronics that cannot be replaced without major intervention.
        Overall the areas of inefficiency have no impact in the endcap muon reconstruction performance.
    }
    \label{fig:csc_performance}
\end{figure}

The on-detector electronics for the entire system has been replaced during LS2, and commissioned and operated in Run 3.
Fig.\,\ref{fig:csc_performance} shows the trigger primitive efficiency of the CSC system during the proton-proton collisions in Run 3 \cite{ref:csc_efficiency}:
most of the system has an efficiency close to 100\%, with local inefficiency areas due to known electronics issues that do not impact the endcap muon trigger performance.

\section{RPC upgrades}

The RPC system instruments the CMS barrel and endcaps, complementing each DT and CSC station except for the four forward CSC stations ME1/1, ME2/1, ME3/1 and ME4/1.
The legacy RPC technology consists of double-gap detectors with bakelite electrodes with \SI{1.5}{\nano\s} time resolution and \SI{1.7}{\milli\radian} space resolution in the $\phi$ coordinate;
the Phase-2 upgrades of the RPC system include the replacement of its legacy link system and the addition of two new forward stations.

The legacy readout electronics in the experimental area consists of a front-end board (FEB) sending data via copper lines to the link board, while the slow control is managed by the control board.
The link system requires an upgrade to replace the ASICs used in the link boards, whose maintenance is not possible since they are unsupported by modern systems:
the new link boards include a Xilinx Kintex-7 FPGA, which in turn implements a TDC to exploit the timing information of the RPC signal with a granularity of \SI{2.5}{\nano\s}, ten times higher than the legacy electronics.
The availability of timing information will be useful -- especially for the new RPC stations -- to reject background and improve the muon segment resolution.

In 2023, the design of the new link board has been validated at the CERN High energy AcceleRator Mixed field/facility (CHARM) \cite{ref:charm} to determine its radiation hardness at the HL-LHC expected doses, allowing the launch of the board pre-production.

\subsection{RPC longevity}

To determine the reliability of the RPC detectors to the HL-LHC background dose, ageing studies are ongoing at the GIF++ since 2016 with the goal of reaching an integrated charge equal to 3 times the expected charge of ten years of Phase-2 operations, equal to approximately \SI{0.8}{\coulomb/\centi\m\squared}.
In 2022, an efficiency measurement performed with SPS muons on an RE2/2 detector after irradiation by 97\% of the total HL-LHC integrated charge has shown no observable variation in the plateau efficiency of the detector;
however, a shift of the plateau HV point by \SI{380}{\volt} was observed, due to the increased resistivity of the HPL electrodes after irradiation.
No efficiency drop was observed under irradiation up to a background particle rate of \SI{600}{\Hz/\centi\m\squared}.

\subsection{New RPC stations}

Two new RPC stations, RE3/1 and RE4/1, will be installed between the Run 3 technical stops and the LHC Long Shutdown 3 (LS3, foreseen from 2026 to 2028);
they will occupy the first ring of, respectively, the third and fourth endcap stations, covering the pseudorapidity region $1.8 < |\eta| < 2.4$ \cite{ref:irpc};
This extension will allow recovering high single-muon trigger efficiency especially in the region $2.2 < |\eta| < 2.4$.

For both endcaps, each of the two stations will be made of 18 RPCs with 20° aperture.
To sustain the expected background particle rate up to \SI{700}{\Hz/\centi\m\squared} in the HL-LHC, the RPC detector design for the new stations has been upgraded with respect to the legacy technology:
the improved RPCs (iRPCs) will be built with smaller gaps (\SI{1.4}{\milli\m} instead of the previous \SI{2}{\milli\m}) and lower resistivity HPL electrodes (up to \SI{3e10}{\ohm\centi\m});
the front-end electronics, based on the Petiroc ASIC \cite{ref:petiroc}, will allow lower noise to apply charge thresholds (on average \SI{50}{\femto\coulomb}) of a factor 3 lower than the legacy RPCs and lower the detector HV point.
Coarse position information in the $\eta$ direction will be available by measuring the delay in the signal arrival times at the two ends of the readout strips.

\begin{figure}[tb]
    \centering
    \includegraphics[width=\textwidth]{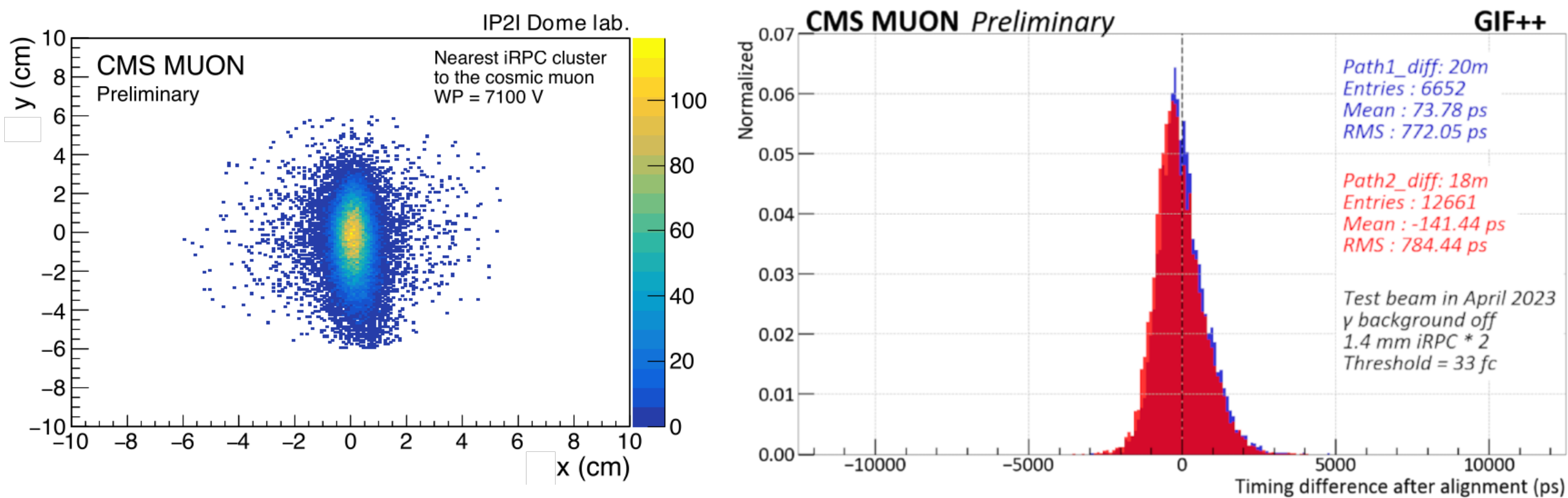}
    \caption{
        (a) Residuals of iRPC measured with cosmic rays at the IP2I Lyon laboratory.
        (b) Arrival time distribution in coincidence of two iRPCs with high-energy muons at the CERN GIF++ \cite{ref:rpc_timing}.
    }
    \label{fig:rpc_performance}
\end{figure}

The performance of the RPC detectors with final electronics has been measured in 2023 with cosmic rays at the IP2I laboratory in Lyon and with muon beams at the CERN GIF++ \cite{ref:rpc_timing}:
the space resolution measured is \SI{1.7}{\centi\m} in the eta direction and \SI{0.4}{\centi\m} in the phi direction (Fig.\,\ref{fig:rpc_performance}a), while the time resolution is about \SI{550}{\pico\second} (Fig.\,\ref{fig:rpc_performance}b).
The mass production of the iRPCs for RE3/1 and RE4/1 has started in 2022 at CERN and in Ghent and is expected to continue in 2024 with the final validation of the production detectors with cosmic rays at CERN.

\section{GEM upgrade}

The Phase-2 GEM upgrade consists of the addition of three stations of triple-GEM detectors in each of the two endcaps to complement the forward CSC stations ME1/1 and ME2/1 and extend the muon system coverage in the very forward region \cite{ref:gem_tdr}.

\subsection{GE1/1 system operations}

The three GEM systems have different schedules, with the first one, GE1/1, being an early Phase-2 upgrade.
GE1/1 has been installed in CMS in LS2 and complements the ME1/1 station in the pseudorapidity region $1.5 < |\eta| < 2.0$;
the higher redundancy and better momentum resolution provided by the GEM hits included in the ME1/1 muon segment reconstruction is expected to allow a reduction in the Level-1 muon trigger rate by a factor 10.

Each of the two GE1/1 endcaps is made of 36 trapezoidal ``super-chambers'' of 10° aperture;
in turn, each superchamber is made of two stacked triple-GEM detectors.
The commissioning of GE1/1 has been carried out during the CMS cosmic runs in LS2 and with pp collisions in Run 3;
the GE1/1 system has recorded collision data with pp and heavy-ion collisions up to 2023 with efficiency larger than 97\% in most of the readout segments.
The stability of the system to discharges under irradiation by the collision background has been improved by tuning the working point of each detector according to its discharge rate and plateau efficiency.

\subsection{GE2/1 detector performance}

The GE2/1 system is expected to be installed in LS3 and during the subsequent technical stops to complement the ME2/1 station in the pseudorapidity region $1.6 < |\eta| < 2.4$.
The GE2/1 detector layout is similar to GE1/1, with the difference that a single GE2/1 chamber has a 20° aperture and is divided along the eta direction in four independent modules for easier manufacturing.
The detector design has been optimized to incorporate the results of the operational experience of GE1/1 by ensuring better planarity in the gas gaps with pillars, segmenting the first and second GEM foils in their bottom electrode to reduce the discharge energy and providing redundant grounding of the readout connectors to lower the electronics noise.

The performance of a GE2/1 detector module has been measured in two test beam campaigns at CERN SPS in 2021 and 2022, reporting excellent local efficiency (close to 100\% except for the areas occupied by the GEM foil segmentation, Fig.\,\ref{fig:ge21_me0}a) and a space resolution of \SI{346+-6}{\micro\radian}, in agreement with its design requirement \cite{ref:gem_tb}.

The mass production of the GE2/1 detectors, started in 2021 in five construction sites, has entered a new stage in 2023 with the final detector validation with cosmic rays at CERN and is expected to continue throughout 2024 until the start of the ME0 production.

\subsection{ME0 design optimization}

The ME0 system is expected to be installed in LS3 in the pseudorapidity region $2.15 < |\eta| < 2.8$, complementing the other GEM and CSC stations in the overlap region and extending the CMS muon system acceptance in the region $2.4 < |\eta| < 2.8$.
In order to reconstruct muon segments in standalone in the region so far not instrumented by any other stations, ME0 will be made of 18 ``stacks'' of triple-GEM detectors per endcap.

\begin{figure}[tb]
	\centering
	\includegraphics[width=\textwidth]{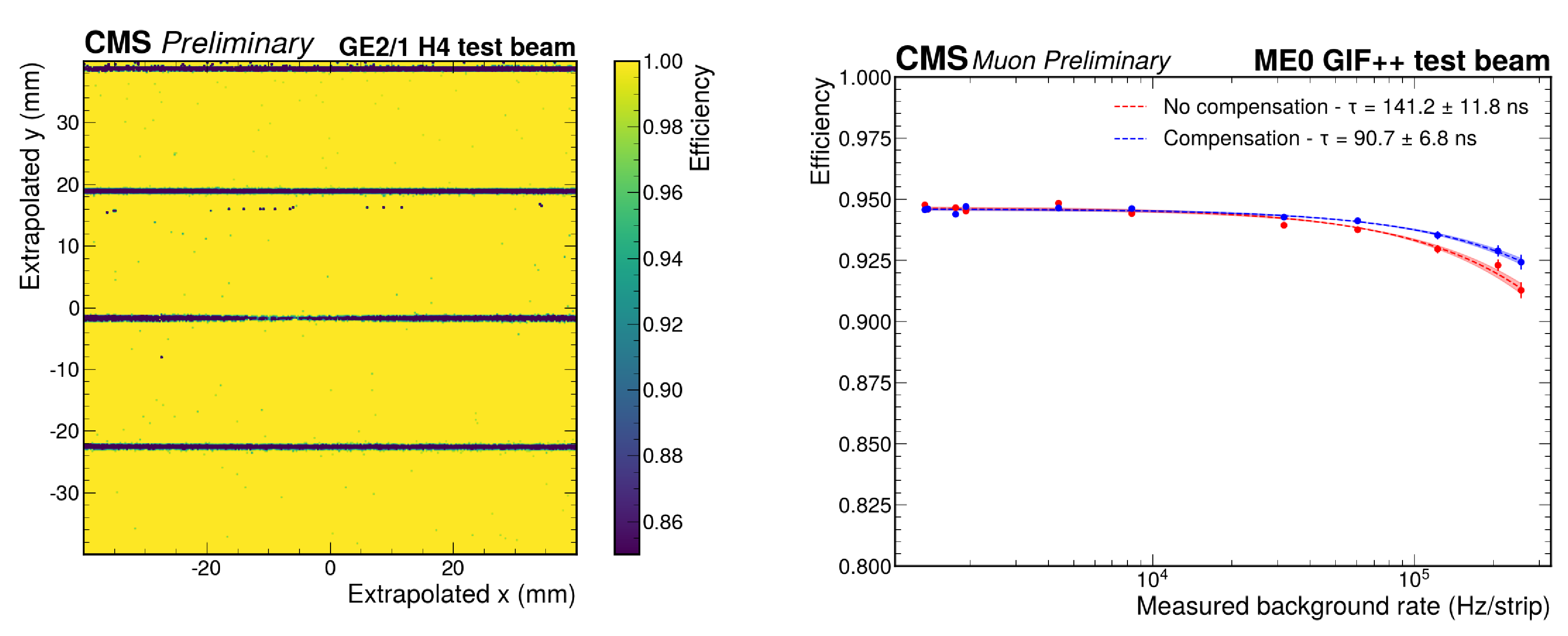}
    \caption{
        (a) Efficiency map of a GE2/1 detector measured at test beam with 80 GeV/c muons.
        (b) Efficiency of an ME0 detector for 80 GeV/c muons measured in test beam at the GIF++ as a function of the background rate.
    }
	\label{fig:ge21_me0}
\end{figure}

While the baseline ME0 technology is the same as GE1/1 and GE2/1, the detector design has required additional design optimizations to sustain the high background rates (up to an average of \SI{150}{\kilo\Hz/\centi\m\squared} in the highest-$\eta$ readout partition) expected at HL-LHC.
After a campaign of rate capability studies in laboratory and in GIF++, the ME0 GEM foils have been segmented in 40 sectors along the $\phi$ direction to limit gain drops under irradiation induced by the foil protection resistors;
a gain recovery routine has been determined to dynamically adjust the detector HV point to the background rate to recover possible gain losses under high irradiation \cite{ref:gem_rate_capability}.
Finally, the rate capability of an ME0 detector with final front-end electronics has been measured in a test beam with high-energy muons and gamma background irradiation at the GIF++;
the results (Fig.\,\ref{fig:ge21_me0}b) show an efficiency drop of 2.5\% at \SI{200}{\kilo\Hz}/strip for a single detector layer due to the dead time of the front-end electronics, which is expected to be mitigated by the high redundancy of the ME0 system (six detectors for each stack).

Ageing studies, performed on an ME0 with an X-ray gun up to an integrated charge of \SI{8}{\coulomb/\centi\m\squared} -- equal to the expected value after 10 years of HL-LHC operations --, have shown no degradation of the detector gain \cite{ref:gem_ageing}.

The performance of full stacks of six triple-GEM detectors for ME0 will be evaluated with dedicated studies in 2024;
the ME0 mass production is expected to start in fall 2024.

\section{Conclusions}

The Phase-2 upgrade of the CMS muon spectrometer involves the electronics upgrade of the three existing detector systems (DTs, CSCs and RPCs) to handle the HL-LHC doses, background and collision rates and the installation of new RPC and GEM stations in the endcap forward region.
The DT electronics upgrade consists mainly of the replacement of the front-end electronics with the OBDT board;
its improved performance has been demonstrated in a Run 3 slice test with a measurement of the muon segment timing -- between 2 and \SI{3}{\nano\s} -- in comparison with the legacy electronics.
The CSC on-detector electronics has been installed during LS2 and operated during Run 3 with good overall efficiency, larger than 95\% for most detectors.
The electronics pre-production of the new link and control boards for the new RPC link system has started in 2023.

The new RPC stations will use an improved RPC (iRPC) technology for higher rate capability and better space and time resolution;
the iRPC performance has been demonstrated in laboratory cosmic measurements and in test beam at the GIF++ and found compliant with its design specifications, with a space resolution of \SI{4}{\milli\m} in the $\phi$ direction and a time resolution of \SI{554}{\pico\second}.
Production of the new RPC detectors is ongoing at CERN and Ghent.

Out of the three GEM stations, GE1/1 has been installed in LS2 and its operation has been demonstrated successfully with good offline performance in Run 3.
The detector performance for the GE2/1 and ME0 stations has been measured in test beams and found in agreement with its design requirements, with an efficiency higher than 99\% and a space resolution of \SI{346+-6}{\micro\radian} for GE2/1 and \SI{235+-2}{\micro\radian} for ME0;
the ME0 detector design has been optimized to handle the high background rates (up to \SI{150}{\kilo\Hz/\centi\m\squared}) and its production is expected to start in 2024.

\end{document}